\documentclass[showpacs,aps,pra,twocolumn]{revtex4}
\usepackage{amsmath,graphicx,bbm,mathrsfs,amssymb,pst-all,bm,color}

\input{tcilatex}
\begin{document}

\title{Efficient quantum algorithm for preparing molecular-system-like
states on a quantum computer}
\author{Hefeng Wang, S. Ashhab and Franco Nori}
\affiliation{Advanced Science Institute, The Institute of Physical and Chemical
Research~(RIKEN), Wako-shi, Saitama 351-0198, Japan\\
and Department of Physics, The University of Michigan, Ann Arbor, Michigan
48109-1040, USA}

\begin{abstract}
We present an efficient quantum algorithm for preparing a pure state on a
quantum computer, where the quantum state corresponds to that of a molecular
system with a given number $m$ of electrons occupying a given number $n$ of
spin orbitals. Each spin orbital is mapped to a qubit: the states $\vert 1
\rangle$ and $\vert 0 \rangle$ of the qubit represent, respectively, whether
the spin orbital is occupied by an electron or not. To prepare a general
state in the full Hilbert space of $n$ qubits, which is of dimension $2^{n}$%
, $O(2^{n})$ controlled-NOT gates are needed, i.e., the number of gates
scales \emph{exponentially} with the number of qubits. We make use of the
fact that the state to be prepared lies in a smaller Hilbert space, and we
find an algorithm that requires at most $O(2^{m+1} n^{m}/{m!})$ gates, i.e.,
scales \emph{polynomially} with the number of qubits $n$, provided $n\gg m$.
The algorithm is simulated numerically for the cases of the hydrogen
molecule and the water molecule. The numerical simulations show that when
additional symmetries of the system are considered, the number of gates to
prepare the state can be drastically reduced, in the examples considered in
this paper, by several orders of magnitude, from the above estimate.

\noindent
\end{abstract}

\maketitle

\section{introduction}

Simulating quantum systems on a classical computer is a hard problem. The
size of the Hilbert space of the simulated system increases exponentially
with the system size. For example, in quantum chemistry, the full
configuration interaction~(FCI) method diagonalizes the molecular
Hamiltonian to provide solutions to the electronic structure problem. The
resource requirement for performing the FCI calculation scales exponentially
with the size of the system~\cite{helgaker}. Therefore, it is restricted to
the treatment of small diatomic and triatomic systems~\cite{LJ}. Feynman~%
\cite{fey} observed that simulating a quantum system might be more efficient
on a quantum computer than on a classical computer. Further work has born
out of this early suggestion~\cite{lloyd, zalka, abrams, lidarwang, man, aa,
whf, guzikdyn, nori, bul, you}.

The implementation of a quantum simulation algorithm requires a mapping from
the system wave function to the state of the qubits. One possible mapping is
through the Jordan-Wigner transformation~(JWT)~\cite{JWT}. Each spin orbital
is mapped to a qubit: the states $\vert 1\rangle $ and $\vert 0\rangle $ of
the qubit represent, respectively, whether the spin orbital is occupied by
an electron or not. The simulated system, and therefore the quantum
computer, can be in a quantum superposition of different configurations. In
quantum chemistry, this quantum superposition is known as the configuration
state function~(CSF). There are other possible mappings between the
simulated system and the quantum computer~\cite{aa, whf}. For example, one
can make use of the fact that the Hilbert space of interest, i.e., one with $%
m $ electrons occupying $n$ spin orbitals, is of dimension $\left(
\begin{array}{c}
n \\
m%
\end{array}%
\right) $, and use the smallest number $n^{\prime }$ of qubits that
satisfies the condition $2^{n^{\prime }}\geq \left(
\begin{array}{c}
n \\
m%
\end{array}%
\right) $. The Hilbert space of the quantum computer would then be large
enough to describe any quantum state of the simulated system. However, since
these compact mappings complicate the implementation of the simulation, we
only consider the simple mapping explained above.

Preparing a general pure state in the full Hilbert space of $n$ qubits,
which is of dimension $2^{n}$, has been studied by several groups. Shende
and Markov~\cite{shende} and M\"{o}tt\"{o}nen \emph{et}~\emph{al.}~\cite%
{mott} showed that preparing a generic $n$-qubit pure state from $|0\rangle
^{\otimes n}$ requires $O(2^{n})$ controlled-NOT~(CNOT) gates, which gives
an exponential scaling for the number of CNOT gates; Bergholm \emph{et}.~%
\emph{al.}~\cite{berg} gave an upper bound for the number of gates required
for transforming an arbitrary state $|a\rangle $ to an arbitrary state $%
|b\rangle $: ($2^{n+1}-2n-2$) CNOT gates and ($2^{n+1}-n-2$) one-qubit
gates. The gate count is halved if $|a\rangle $ or $|b\rangle $ is one of
the basis configuration states. Another quantum algorithm for the
preparation of an arbitrary pure state with fidelity arbitrarily close to
one was suggested by Soklakov and Schack~\cite{sok}, which is based on
Grover's quantum search algorithm.

In this paper, we study the preparation of a pure state in the
second-quantized representation that represents the quantum state of $m$
electrons distributed among $n$ spin orbitals. The state to be prepared lies
in the combinatorial space of dimension $\left(
\begin{array}{c}
n \\
m%
\end{array}%
\right) $, which is a subspace of the full Hilbert space of $n$ qubits.
Ortiz \emph{et}~\emph{al.}~\cite{ortiz} studied a similar problem, and they
gave an algorithm for preparing a state that is composed of $N$
configurations, where $N$ is a finite and small number. Their algorithm
scales as $O(N^{2}n^{2})$, where $n$ is the number of the qubits. In this
paper we present an efficient recursive algorithm that gives specific
quantum circuit for preparing a pure state, with polynomial scaling of the
number of CNOT gates in terms of the number of the qubits. We have
numerically simulated the state preparation algorithm to prepare the
electronic states of the hydrogen and the water molecules. The results show
that the number of CNOT gates can be reduced by up to $3$ orders of
magnitude from the upper bound we derive.

The structure of this work is as follows. In Sec.~\ref{SSsec} we discuss
mapping the Fock space of the system onto the Hilbert space of the qubits.
In Sec.~\ref{SPsec} we present a recursive algorithm for state preparation.
Preparing a general pure state in the one-electron systems is discussed in
detail. Then based on the results for the one-electron system, a recursive
approach for preparing a general state of an $m$-electron system is
presented. In Sec.~I\negthinspace V, we discuss the simplification of the
two-fold controlled unitary operations in the algorithm. In Sec.~V, we
analyze the scaling of the algorithm. In Sec.~V\negthinspace I, we apply the
algorithm to prepare the electronic states of the hydrogen and the water
molecules. We close with a conclusion section.

\section{Fock space of molecular systems and the Jordan-Wigner Transformation%
}

\label{SSsec}

The wave function of a $m$-electron, $n$-spin-orbital system is a linear
combination of Slater determinants. In the formalism of second quantization
it is written as:
\begin{equation}
|\Psi \rangle =\sum_{\substack{ i_{1},\text{ }\cdots ,\text{ }i_{n}\in
\{0,1\}  \\ i_{1}\text{ }+\text{ }\cdots \text{ }+\text{ }i_{n}=m}}%
k_{i_{1},\cdots ,i_{n}}(a_{1}^{\dagger })^{i_{1}}\cdots (a_{n}^{\dagger
})^{i_{n}}|\Omega \rangle ,
\end{equation}%
where $|\Omega \rangle $ represents the vacuum state with no electrons. Each
set $\{i_{1},\cdots ,i_{n}\}$ represents an electron configuration of the
system, and these states are used as the basis vectors of the configuration
space of the system. The fermion creation and annihilation operators $%
a_{j}^{\dag }$ and $a_{j}$ satisfy the canonical anticommutation relations $%
\{a_{i},a_{j}\}=0; \{a_{i},a_{j}^{\dag }\}=\delta _{ij}$. $a_{j}^{\dag }$ ($%
a_{j}$) creates~(annihilates) a fermion on the $j$th spin orbital.

The fermion algebra is isomorphic to the standard quantum computing~(QC)
model~(or Pauli) algebra. The isomorphism is established through the
Jordan-Wigner transformation~\cite{JWT}. The JWT maps a fermion state to a
one-dimensional standard QC state and vice versa. The creation and
annihilation operators for the fermion state are mapped to the Pauli
operators through the JWT,
\begin{equation}
a_{j}\rightarrow \lbrack \Pi _{i=1}^{j-1}(-\sigma _{z}^{i})]\sigma _{-}^{j},
\end{equation}%
\begin{equation}
a_{j}^{\dag }\rightarrow \lbrack \Pi _{i=1}^{j-1}(-\sigma _{z}^{i})]\sigma
_{+}^{j},
\end{equation}%
where $\sigma _{z}$ is the Pauli matrix defined as:%
\begin{equation}
\sigma _{z}=\left(
\begin{array}{cc}
1 & 0 \\
0 & -1%
\end{array}%
\right).
\end{equation}%
The operators $\sigma _{-}$ and $\sigma _{+}$ are the Pauli lowering and
raising operators defined as:

\begin{equation}
\sigma _{-}=\frac{1}{2}(\sigma _{x}-i\sigma _{y})=\left(
\begin{array}{cc}
0 & 0 \\
1 & 0%
\end{array}%
\right) ,
\end{equation}

\begin{equation}
\sigma _{+}=\frac{1}{2}(\sigma _{x}+i\sigma _{y})=\left(
\begin{array}{cc}
0 & 1 \\
0 & 0%
\end{array}%
\right) .
\end{equation}%
They satisfy the following conditions:%
\begin{equation}
\sigma _{+}|0\rangle =0,\qquad \sigma _{-}|0\rangle =|1\rangle ,
\end{equation}%
and
\begin{equation}
\sigma _{+}|1\rangle =|0\rangle ,\text{ \ \ \ \ }\sigma _{-}|1\rangle =0,
\end{equation}%
where

\begin{equation}
|0\rangle =\left(
\begin{array}{c}
1 \\
0%
\end{array}%
\right) ,\qquad |1\rangle =\left(
\begin{array}{c}
0 \\
1%
\end{array}%
\right) .
\end{equation}

\section{recursive algorithm for state preparation}

\label{SPsec}

State preparation means starting from a standard initial state on a quantum
computer and, by applying a unitary operation, transforming it to a given
target state. In this paper, the target state in general is an entangled
state mapped from a state in the Fock space of the $m$-electron, $n$%
-spin-orbital system. The unitary operation can be decomposed into a
sequence of elementary quantum gates.

Our approach for state preparation is as follows: instead of starting from
the initial state and transforming it to the target state, we start from the
target state and transform it to the initial state. We can obtain the gate
sequence for preparing the target state by \emph{inverting} the quantum
circuit sequence, since any unitary operation is reversible. A recursive
procedure is applied in our \textquotedblleft reverse engineering$"$
approach. We first solve the problem of state preparation for one-electron
system, then based on this, the general state preparation for $m$-electron
system is solved.

We set the initial state of an $n$-qubit system to be $|0\rangle ^{\otimes
n} $ and define the target state as:
\begin{equation}
|\Psi _{T}\rangle =\sum_{i=1}^{N}k_{i}|\psi _{i}\rangle ,
\end{equation}%
where $N$ is the dimension of the configuration space and
\begin{equation}
\sum_{i=1}^{N}|k_{i}|^{2}=1,
\end{equation}%
where $|\psi _{i}\rangle $ is a single configuration that describes a
distribution of $m$ electrons on $n$ spin orbitals. The states $\{|\psi
_{i}\rangle \}$ are the basis vectors of the configuration space. For
example, the state $\frac{1}{\sqrt{3}}|001\rangle +\frac{1}{\sqrt{6}}%
|010\rangle +\frac{1}{\sqrt{2}}|100\rangle $ is composed of three
configurations in the state space of a one-electron, three-spin-orbital
system.

\subsection{Gate library}

The gate library used in our approach contains the CNOT gate and the single
qubit gates $C$. Here $C$ represents all unitary single qubit operators in
SU($2$). It can be written in the form:
\begin{equation}
C=u|0\rangle \langle 0|+v|1\rangle \langle 0|-v^{\ast }|0\rangle \langle
1|+u^{\ast }|1\rangle \langle 1|=\left(
\begin{array}{cc}
u & -v^{\ast } \\
v & u^{\ast }%
\end{array}%
\right)
\end{equation}%
where $u$ and $v$ are complex numbers and
\begin{equation}
|u|^{2}+|v|^{2}=1.
\end{equation}%
The Hadamard gate $H$ and the NOT gate $X$ are included in $C$. We define
another class of operators $\widetilde{H}$, which is a sub-class of the
single-qubit operator $C$, such that:
\begin{equation}
\widetilde{H}=C^{\dag }XC=\left(
\begin{array}{cc}
u^{\ast }v+uv^{\ast } & {u^{\ast }}^{2}-{v^{\ast }}^{2} \\
u^{2}-v^{2} & -(u^{\ast }v+uv^{\ast })%
\end{array}%
\right) ,
\end{equation}%
for some $C$. We call $\widetilde{H}$ the generalized Hadamard gate. It has
the same property as the Hadamard gate, $\widetilde{H}\widetilde{H}=I$, and
acts in a similar way as the Hadamard gate. Just like the Hadamard gate
transforms a superposition state $\frac{1}{\sqrt{2}}(|0\rangle +|1\rangle )$
of a single qubit to state $|0\rangle $ and $\frac{1}{\sqrt{2}}(|0\rangle
-|1\rangle )$ to state $|1\rangle $, the $\widetilde{H}$ gate transforms a
given superposition state $a|0\rangle + b|1\rangle $ of a single qubit to
state $|0\rangle $ or $|1\rangle $.

A considerable effort has been made to synthesize two-qubit circuits using
CNOT gates and single-qubit gates~\cite{vidal, vatan, vshende, zhang} and
quantum logic circuits~\cite{vart, mot, vvvshende, tuc}. It has been shown
that to implement a typical two-qubit operator, three CNOT gates are needed.
In this paper we only use two kinds of two-qubit gates: the CNOT gate and
the controlled-$\widetilde{H}$ gate, $C$-$\widetilde{H}$, which as shown in
Fig.~$1$, contains only one CNOT gate. In our algorithm, the $\widetilde{H}$
and $C$-$\widetilde{H}$ gates are used to address the configuration
coefficients and reduce the number of the configurations that span the
target state $\vert \Psi _{T}\rangle $ until it reaches the initial state $%
\vert 0 \rangle ^{\otimes n}$. An example for transforming a three qubit
state to the initial state is shown below.
\begin{figure}[tbp]
\centering
\includegraphics[width=0.45\columnwidth, clip]{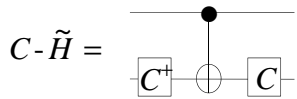}
\caption{Quantum circuit for the controlled-$\widetilde{H}$ gate.}
\label{ch}
\end{figure}

To transform the target state to the initial state, we first disentangle the
target state, then reduce the number of configurations in the target state
by applying $\widetilde{H}$ and $C$-$\widetilde{H}$ gates. A simple example
for a Bell-type state $|\beta \rangle =a|10\rangle +b|01\rangle $ is shown
as follows: to transform this state to the initial state $|00\rangle $,
first we apply the $X^{1}$ gate to the first qubit~(the number $``1"$ on the
superscript indicates that the operation is applied to the first qubit). We
thus obtain $X^{1}|\beta \rangle =a|00\rangle +b|11\rangle $. We then
disentangle this state by applying a CNOT gate: $\mathrm{{CNOT^{1,2}}}$%
~(where on the superscript, the first number represents the control qubit
and the second number represents the target qubit). Therefore we now obtain $%
\mathrm{{CNOT^{1,2}}}\cdot X^{1}|\beta \rangle =a|00\rangle +b|10\rangle
=(a|0\rangle +b|1\rangle )|0\rangle $. Finally by applying a generalized
Hadamard gate $\widetilde{H}^{1}$ to the first qubit, we obtain the initial
state $|00\rangle $. So, $|00\rangle =\widetilde{H}^{1}\cdot \mathrm{{%
CNOT^{1,2}}}\cdot X^{1}|\beta \rangle $. Thus the inverse gate sequence will
transform the initial state $|00\rangle $ to the target state $|\beta
\rangle $. The quantum circuit for this process is shown in Fig.~$2$.
\begin{figure}[tbp]
\centering
\includegraphics[width=0.45\columnwidth, clip]{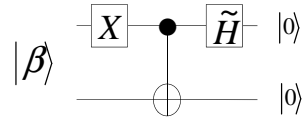}
\caption{Quantum circuit for transforming a Bell-type state $|\protect\beta %
\rangle =a|10\rangle +b|01\rangle )$ to the initial state $|00\rangle $. So,
$|00\rangle =\widetilde{H}^{1}\cdot CNOT\cdot X^{1}\cdot |\protect\beta %
\rangle$.} \label{bell}
\end{figure}

The following is an example that shows how the configuration coefficients
can be addressed and the number of the configurations in the target state
can be reduced through the application of the generalized Hadamard gate $%
\widetilde{H}$ and $C$-$\widetilde{H}$ gate. To transform the state $|\Psi
_{T}\rangle =\frac{1}{\sqrt{3}}|001\rangle +\frac{1}{\sqrt{6}}|010\rangle +%
\frac{1}{\sqrt{2}}|100\rangle $, to the state $|000\rangle =|0\rangle
^{\otimes 3}$, the gate sequence is as follows:
\begin{eqnarray}
&&|\Psi _{T}\rangle \text{ \ }\underrightarrow{\text{ }X^{1}\text{ }}\text{
\ }|1\rangle (\frac{1}{\sqrt{3}}|01\rangle +\frac{1}{\sqrt{6}}|10\rangle )+%
\frac{1}{\sqrt{2}}|0\rangle ^{\otimes 3}  \notag \\
&&\underrightarrow{\text{ }CNOT^{1,2}\text{ }}\text{\ \ }|1\rangle (\frac{1}{%
\sqrt{3}}|11\rangle +\frac{1}{\sqrt{6}}|00\rangle )+\frac{1}{\sqrt{2}}%
|0\rangle ^{\otimes 3}  \notag \\
&&\underrightarrow{\text{ }CNOT^{2,3}\text{ }}\text{ \ }|1\rangle (\frac{1}{%
\sqrt{3}}|10\rangle +\frac{1}{\sqrt{6}}|00\rangle )+\frac{1}{\sqrt{2}}%
|0\rangle ^{\otimes 3}  \notag \\
&&\underrightarrow{\text{ \ \ }C\text{-}{\widetilde{H}}^{1,2}\text{ }}\text{
\ }\frac{1}{\sqrt{2}}|100\rangle +\frac{1}{\sqrt{2}}|0\rangle ^{\otimes 3}
\notag \\
&&\underrightarrow{\text{ \ \ \ \ \ }H^{1}\text{\ \ \ }}\text{ \ }|0\rangle
^{\otimes 3},
\end{eqnarray}%
where in the $C$-$\widetilde{H}$ operation, the $\widetilde{H}$ gate is
determined by solving the following equation:
\begin{equation}
\widetilde{H}\left(
\begin{array}{c}
\frac{1}{\sqrt{6}} \\
\frac{1}{\sqrt{3}}%
\end{array}%
\right) =\left(
\begin{array}{cc}
2uv & u^{2}-v^{2} \\
u^{2}-v^{2} & -2uv%
\end{array}%
\right) \left(
\begin{array}{c}
\frac{1}{\sqrt{6}} \\
\frac{1}{\sqrt{3}}%
\end{array}%
\right) =\frac{1}{\sqrt{2}}\left(
\begin{array}{c}
1 \\
0%
\end{array}%
\right) .
\end{equation}%
We set $u$ and $v$ in the unitary matrix $C$ to be real numbers since the
coefficients in the target state are real. Then one can obtain the unitary
matrix $C$ by solving the above equation; one obtains $u=\frac{1}{6}(3\sqrt{2%
}+2\sqrt{3})\sqrt{3-\sqrt{6}}$ and $v=-\sqrt{1/2-1/\sqrt{6}}$. The quantum
circuit for this procedure is shown in Fig.~$3$.\newline
\begin{figure}[h]
\includegraphics[width=0.7\columnwidth, clip]{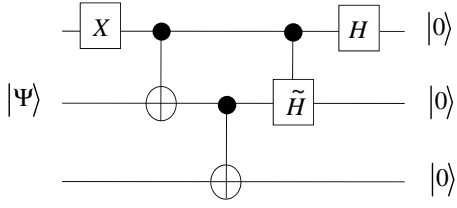}
\caption{Quantum circuit for transforming the three-qubit state
$|\Psi
_{T}\rangle =\frac{1}{\protect\sqrt{3}}|001\rangle +\frac{1}{\protect\sqrt{2}%
}|100\rangle +\frac{1}{\protect\sqrt{6}}|010\rangle $ to the state $%
|0\rangle ^{\otimes 3}=|000\rangle $.}
\label{three}
\end{figure}

\subsection{One-electron system}

Let $|\Psi_{T}(n,1)\rangle $ denote the state wave function of the
one-electron, $n$-spin-orbital system. If we consider all possible
distributions of the electron on the $n$ spin orbitals, the dimension of the
configuration space will be $\left(
\begin{array}{c}
n \\
1%
\end{array}%
\right) =n$. We call this the complete configuration space of the
one-electron system. The target state can be factorized as follows:
\begin{equation}
|\Psi _{T}(n,1)\rangle =c_{0}|0\rangle |\Psi _{T}(n-1,1)\rangle
+c_{1}|1\rangle |0\rangle ^{\otimes (n-1)},
\end{equation}%
where $|\Psi _{T}(n-1,1)\rangle $ is a state wave function for one electron
distributed on $(n-1)$ spin orbitals. The configuration coefficients in
state $|\Psi _{T}(n-1,1)\rangle $ are normalized to $1$. Also, $c_{0}$ and $%
c_{1}$ satisfy $|c_{0}|^{2}+|c_{1}|^{2}=1$. The relative phase between two
states can be addressed by some unitary operations. We define an unitary
operator $Q(n-1,1)$, such that
\begin{equation}
Q(n-1,1)|\Psi _{T}(n-1,1)\rangle =|0\rangle ^{\otimes (n-1)}.
\end{equation}%
The procedure for transforming the target state to the state $|0\rangle
^{\otimes n}$ can be formulated as follows:
\begin{eqnarray}
&&|\Psi _{T}(n,1)\rangle \text{ }\underrightarrow{\text{ }X^{1}\text{ }}%
\text{ }c_{0}|1\rangle |\Psi _{T}(n-1,1)\rangle +c_{1}|0\rangle ^{\otimes
n}\qquad  \notag \\
&&\underrightarrow{\text{ }C\text{-}Q(n-1,1)^{\{1,(2,\cdots ,n)\}}\text{ }}%
\text{ }c_{0}|1\rangle |0\rangle ^{\otimes (n-1)}+c_{1}|0\rangle ^{\otimes
n}\qquad  \notag \\
&&\underrightarrow{\text{ }\widetilde{H}^{1}\text{ }}\text{ }|0\rangle
^{\otimes n},
\end{eqnarray}%
where $C$-$Q(n-1,1)^{\{1,(2,\cdots ,n)\}}$ is a controlled unitary operation
with the first qubit as the control qubit, and qubits $2,\cdots ,n$ are the
target qubit. Here, the $\widetilde{H}$ gate operates on the vector $\left(
\begin{array}{c}
c_{1} \\
c_{0}%
\end{array}%
\right) $ and transforms it to the state $\left(
\begin{array}{c}
1 \\
0%
\end{array}%
\right) =|0\rangle $. Thus the gate $\widetilde{H}$ used here is determined
by solving the following equation:
\begin{equation}
\widetilde{H}\left(
\begin{array}{c}
c_{1} \\
c_{0}%
\end{array}%
\right) =\left(
\begin{array}{cc}
u^{\ast }v+uv^{\ast } & {u^{\ast }}^{2}-{v^{\ast }}^{2} \\
u^{2}-v^{2} & -(u^{\ast }v+uv^{\ast })%
\end{array}%
\right) \left(
\begin{array}{c}
c_{1} \\
c_{0}%
\end{array}%
\right) =\left(
\begin{array}{c}
1 \\
0%
\end{array}%
\right) .
\end{equation}%
The quantum circuit for this procedure is shown in Fig.~$4$.
\begin{figure}[h]
\includegraphics[width=0.8\columnwidth, clip]{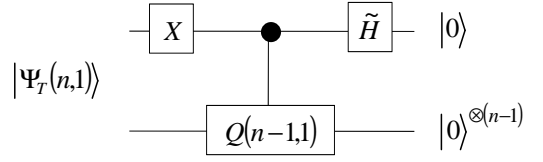}
\caption{Quantum circuit for transforming a general state of the
one-electron system $\vert \Psi _{T}(n,1)\rangle $ to the initial state $%
\vert 0\rangle ^{\otimes n}$.}
\label{1e}
\end{figure}

The unitary operator $Q(n-1,1)$ that transforms the state $\vert \Psi
_{T}(n-1,1)\rangle $ to the state $\vert 0\rangle ^{\otimes (n-1)}$ can be
factorized similarly through the factorization of the state $\vert
\Psi_{T}(n-1,1)\rangle$,
\begin{equation}
\vert \Psi_{T}(n-1,1)\rangle =c_{0}^{\prime }\vert 0\rangle \vert
\Psi_{T}(n-2,1)\rangle +c_{1}^{\prime }\vert 1\rangle \vert 0\rangle
^{\otimes (n-2)},
\end{equation}%
where $\vert c_{0}^{^{\prime }}\vert ^{2}+\vert c_{1}^{\prime }\vert ^{2}=1$
and $\vert \Psi _{T}(n-2,1)\rangle $ is normalized to $1$. An $\widetilde{H}%
^{\prime }$ gate acts on the vector $\left(
\begin{array}{c}
c_{1}^{\prime } \\
c_{0}^{\prime }%
\end{array}%
\right) $ and transforms it to the state $\left(
\begin{array}{c}
1 \\
0%
\end{array}%
\right) =\vert 0\rangle $. The $\widetilde{H}^{^{\prime }}$ gate can be
determined by solving an equation that is similar to Eq.~($20$). We now
define a unitary operator $Q(n-2,1)$, such that
\begin{equation}
Q(n-2,1)\vert \Psi _{T}(n-2,1)\rangle =\vert 0\rangle ^{\otimes (n-2)}.
\end{equation}%
The decomposition procedure above is repeated until the target state reaches
a Bell-type state $\vert \Psi _{T}(2,1)\rangle =c_{0}^{\prime \prime }\vert
01\rangle +c_{1}^{\prime \prime }\vert 10\rangle $, whose disentanglement
procedure has been explained above~(Fig.~$2$). The quantum circuit for this
procedure of decomposing the unitary operator $Q(n-1,1)$ is shown in Fig.~$5$%
.
\begin{figure}[h]
\includegraphics[width=0.8\columnwidth, clip]{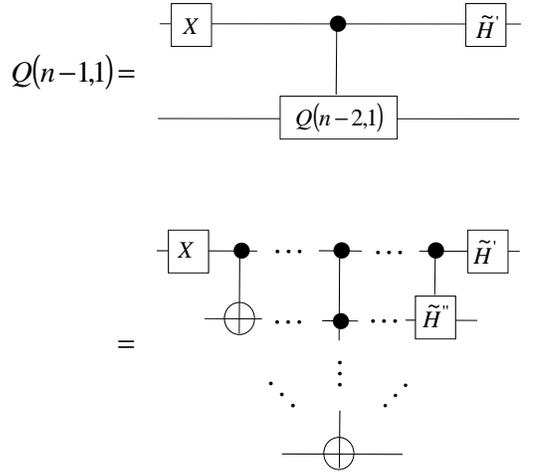}
\caption{Quantum circuit for the decomposition of the unitary operator $%
Q(n-1,1)$.}
\label{du}
\end{figure}

From the derivation above, we can see that the preparation of the state $%
\vert \Psi _{T}(n,1)\rangle $ is reduced to the preparation of the state $%
\vert \Psi _{T}(2,1)\rangle $, which is solved above as shown in Fig.~$2$.
The Toffoli gate, which is also known as the \textquotedblleft
controlled-controlled-not\textquotedblright\ gate, appears when implementing
this recursive procedure. We will discuss the simplification of the Toffoli
gates into CNOT gates in Sec. IV. The scaling of the algorithm in terms of
the number of CNOT and single-qubit gates is analyzed in Sec.~V.

States in the incomplete configuration space of the one-electron system are
easier to prepare than those in the complete configuration space, since the
former can be reduced to states in a complete configuration space of lower
dimension.

\subsection{$m$-electron system}

For the $m$-electron system, it becomes more difficult to prepare the target
state than for the one-electron system. We apply a recursive approach to
solve this problem based on the following facts:

\noindent (1) Any state of the one-electron system can be prepared~[as shown
in Sec.~I\!I\!I.~B.]

\noindent (2) An arbitrary state $\vert \Psi _{T}(p,p-1)\rangle $ can be
prepared. [By applying $X^{\otimes p}$ gates, the state is transformed to $%
\vert \Psi _{T}(p,1)\rangle $, which is a one-electron state.]

We now define a unitary operator $U(n,m)$ such that:
\begin{equation}
U(n,m)\vert \Psi _{T}(n,m)\rangle =\vert 0\rangle ^{\otimes n}.
\end{equation}%
A quantum circuit for $U(n,m)$ can be obtained through the decomposition of
the state $\vert \Psi _{T}(n,m)\rangle $ as follows.

Any arbitrary target state~[Eq.~($10$)] can be rewritten in the following
way:
\begin{equation}
\vert \Psi _{T}\rangle =\sum_{i=1}^{N_{0}}k_{0,i}\vert 0\rangle \vert \psi
_{i}\rangle +\sum_{j=1}^{N_{1}}k_{1,j}\vert 1\rangle \vert \psi _{j}\rangle ,
\end{equation}%
where the state is divided in two parts: one of them lies in the subspace
where the first qubit is in state $\vert 0\rangle $~(we call this the $\vert
0\rangle $ subspace); the other part lies in the subspace where the first
qubit is in state $\vert 1\rangle $~(we call this the $\vert 1\rangle $
subspace). Also, $N_{0}$ is the dimension of the $\vert 0\rangle $ subspace
and $N_{1}$ is the dimension of the $\vert 1\rangle $ subspace. The target
state can be factorized as follows:
\begin{equation}
\vert \Psi _{T}(n,m)\rangle =c_{0}\vert 0\rangle \vert \Psi
_{T}(n-1,m)\rangle +c_{1}\vert 1\rangle \vert \Psi _{T}(n-1,m-1)\rangle .
\end{equation}%
The coefficients $c_{0}$ and $c_{1}$ are calculated as:
\begin{equation}
\vert c_{0}\vert ^{2}=\sum_{i=1}^{N_{0}}\vert k_{0,i}\vert ^{2},\qquad \vert
c_{1}\vert ^{2}=\sum_{j=1}^{N_{1}}\vert k_{1,j}\vert ^{2},
\end{equation}%
where $\vert c_{0}\vert ^{2}+\vert c_{1}\vert ^{2}=1$. We now define unitary
operators $U(n-1,m)$ and $U(n-1,m-1)$ such that
\begin{equation}
U(n-1,m)\vert \Psi _{T}(n-1,m)\rangle =\vert 0\rangle ^{\otimes (n-1)},
\end{equation}%
\begin{equation}
U(n-1,m-1)\vert \Psi _{T}(n-1,m-1)\rangle =\vert 0\rangle ^{\otimes (n-1)}.
\end{equation}%
The relative phase between two configurations can be included in these
unitary operators. Then the target state can be transformed to the initial
state $\vert 0\rangle ^{\otimes n}$ in the following way:
\begin{eqnarray}
&&\vert \Psi _{T}(n,m)\rangle \text{ }\underrightarrow{\text{ }X^{1}\text{ }}%
\text{\ }c_{0}\vert 1\rangle \vert \Psi _{T}(n-1,m)\rangle  \notag \\
&&+c_{1}\vert 0\rangle \vert \Psi _{T}(n-1,m-1)\rangle  \notag \\
&&\underrightarrow{\text{ \ }U(n-1,m-1)\text{ \ }}  \notag \\
&&c_{0}\vert 1\rangle U(n-1,m-1)\vert \Psi _{T}(n-1,m)\rangle +c_{1}\vert
0\rangle ^{\otimes n}  \notag \\
&&\underrightarrow{\text{ \ }{C}\text{{-}}{U(n-1,m-1)^{-1}}^{\{1,(2,\cdots
,n)\}}\text{ \ }}\text{ }  \notag \\
&&c_{0}\vert 1\rangle \vert \Psi _{T}(n-1,m)\rangle +c_{1}\vert 0\rangle
^{\otimes n}  \notag \\
&&\underrightarrow{\text{ \ }C\text{-}U(n-1,m)^{\{1,(2,\cdots ,n)\}}\text{ \
}}  \notag \\
&&c_{0}\vert 1\rangle \vert 0\rangle ^{\otimes (n-1)}+c_{1}\vert 0\rangle
^{\otimes n}  \notag \\
&&\underrightarrow{\text{ \ }\widetilde{H}^{1}\text{ \ }}\text{ }\vert
0\rangle ^{\otimes n}.
\end{eqnarray}%
The quantum circuit for this procedure is shown in Fig.~$6$.
\begin{figure}[h]
\includegraphics[width=0.85\columnwidth, clip]{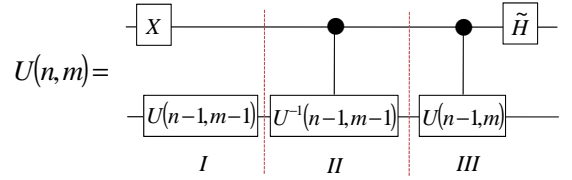}
\caption{Quantum circuit for transforming a general state $\vert \Psi
_{T}(n,m)\rangle $ of the $m$-electron system to the initial state $\vert
0\rangle ^{\otimes {n}}$.}
\label{me}
\end{figure}

The unitary operators $U(n-1,m)$ and $U(n-1,m-1)$ that transform the states $%
|\Psi _{T}(n-1,m)\rangle $ and $|\Psi _{T}(n-1,m-1)\rangle $ to the state $%
|0\rangle ^{\otimes (n-1)}$ can be decomposed through the decomposition of
the states $|\Psi _{T}(n-1,m)\rangle $ and $|\Psi _{T}(n-1,m-1)\rangle $.
For example, for the operator $U(n-1,m-1)$, we have:
\begin{eqnarray}
|\Psi _{T}(n-1,m-1)\rangle &=&c_{0}^{\prime }|0\rangle |\Psi
_{T}(n-2,m-1)\rangle  \notag \\
&&+c_{1}^{\prime }|1\rangle |\Psi _{T}(n-2,m-2)\rangle .
\end{eqnarray}%
Define unitary operators $U(n-2,m-1)$ and $U(n-2,m-2)$ such that
\begin{equation}
U(n-2,m-1)|\Psi _{T}(n-2,m-1)\rangle =|0\rangle ^{\otimes (n-2)},
\end{equation}%
\begin{equation}
U(n-2,m-2)|\Psi _{T}(n-2,m-2)\rangle =|0\rangle ^{\otimes (n-2)}.
\end{equation}%
The quantum circuit for $U(n-1,m-1)$ is similar to the circuit in Fig.~$6$,
where $n$ and $m$ in the unitary operators on the bottom line of the circuit
are replaced by $(n-1)$ and $(m-1)$, respectively.

One can repeat this procedure until reaches a one-electron state $\vert \Psi
_{T}(p,1)\rangle $, where $2\leq p\leq n$, or a $\vert \Psi
_{T}(p,p-1)\rangle $ state, which is solved as we discussed above. In each
iteration, the unitary operators can be decomposed into the circuits that
have the same structure as shown in Fig.~$6$. Such that the problem of
transforming a general $m$-electron, $n$-spin-orbital state $\vert \Psi
_{T}(n,m)\rangle $ to the initial state $\vert 0\rangle ^{\otimes n}$ can be
solved. This algorithm also works for systems of different numbers of the
electrons. The simplification of the twofold controlled unitary operations
and the scaling of the algorithm are discussed in Secs. I\!V and V.

\section{Simplification of the two-fold controlled gates}

\label{TFsec}

Because controlled gates appear inside the controlled operations in
Sec.~I\!I\!I, the twofold controlled gates appear in the implementation of
the recursive procedure for state preparation. For the simplest twofold
controlled gates, the Toffoli gate~\cite{toff}, which is also known as the
\textquotedblleft controlled-controlled-not\textquotedblright\ gate, in
general, six CNOT gates are needed for implementing a three-qubit Toffoli
gate~\cite{barenco}. However, if the input state for the quantum circuit is
known, then the operation of a given unitary operator on all orthogonal
states is immaterial. In such cases the unitary operator is said to be
incompletely specified~\cite{shende}. Thus, there exists more than one
quantum circuit for performing such an operation. So, one can select a
simpler circuit in order to reduce the number of gates. Based on this, we
will now simplify the two-fold controlled gates in the algorithm.

In our algorithm, the problem of preparing an $n$-qubit state is reduced to
the problem of preparing some simpler states~[i.e., for one-electron system,
the problem is reduced to the preparation of a Bell-type state; for a
general $m$-electron system, the problem is reduced to the preparation of
states $\vert \Psi _{T}(p,1)\rangle $ and $\vert \Psi_{T}(p,p-1)\rangle $].
We will show that the twofold controlled gates that appear in each step of
this recursive procedure can be simplified to onefold controlled gates by
just turning off the operation of the first control qubit. The recursive
approach for constructing the quantum circuit leads to another advantage:
for the controlled operations, it provides a way that makes the shortest
distance between the control qubit and the target qubit since the controlled
operations are simplified in each step while expanded to one more qubit.

For the one-electron system, plugging the circuit for $Q(n-1,1)$ shown in
Fig.~$5$ into Fig.~$4$, the quantum circuit for transforming the target
state $\vert \Psi _{T}(n,1)\rangle $ to the state $\vert 0\rangle ^{\otimes
n}$ is shown in the top part of Fig.~$7$. The quantum circuit has the
following structure: from right to left, the input state is $\vert 0\rangle
^{\otimes n} $; an $\widetilde{H}$ gate operates on the first qubit and
followed by a $C$-$\widetilde{H}^{^{\prime }}$ gate. After the application
of these two gates, the basis vectors that span the state on the two control
qubits are $\{\vert 00\rangle $, $\vert 10\rangle $, $\vert 11\rangle \}$.
The basis vector $\vert 01\rangle $ does not appear, such that the two-fold
controlled gates following these operations are incompletely specified. From
right to left, taking $\vert 0\rangle ^{\otimes n}$ as the input state, the
basis vectors that appear in the state of the qubits change as follows:
\begin{widetext}
\begin{equation}
\vert  0\rangle ^{\otimes n}\text{ }\underrightarrow{\widetilde{H}^{1}}\text{ }%
\left \{
\begin{array}{c}
\vert  0\rangle \vert  0\rangle ^{\otimes (n-1)} \\
\vert  1\rangle \vert  0\rangle ^{\otimes (n-1)}%
\end{array}%
\right \} \underrightarrow{{C \text{{-}} \widetilde{H}^{^{\prime
}}}^{1,2}}\left \{
\begin{array}{c}
\vert  0\rangle \vert  0\rangle \vert  0\rangle ^{\otimes (n-2)} \\
\vert  1\rangle \vert  0\rangle \vert  0\rangle ^{\otimes (n-2)} \\
\vert  1\rangle \vert  1\rangle \vert  0\rangle ^{\otimes (n-2)}%
\end{array}%
\right \} \underrightarrow{C \text{{-}}
{Q(n-2,1)^{-1}}^{(1,2),(3,\cdots ,n)}}\left \{
\begin{array}{c}
\vert  0\rangle \vert  0\rangle \vert  0\rangle ^{\otimes (n-2)} \\
\vert  1\rangle \vert  0\rangle \vert  0\rangle ^{\otimes (n-2)} \\
\vert  1\rangle \vert  1\rangle Q(n-2,1)^{-1}\vert  0\rangle ^{\otimes (n-2)}%
\end{array}%
\right \}.
\end{equation}
\end{widetext}The twofold controlled unitary operator $C$-${Q^{-1}(n-2,1)}%
^{(1,2),(3,\cdots ,n)}$ is incompletely specified. Here the use of $C$-${%
Q^{-1}(n-2,1)}^{(1,2),(3,\cdots ,n)}$ instead of $C$-$Q(n-2,1)^{(1,2),(3,%
\cdots ,n)}$ is because we let the gates operate to the right. As a result,
it can be simplified to $C$-${Q^{-1}(n-2,1)}^{2,(3,\cdots ,n)}$. This
simplification can be repeated in each recursive step until the state
reaches a Bell-type state $\vert \Psi _{T}(2,1)\rangle $. The simplified
quantum circuit is shown in the bottom part of Fig.~$7$.
\begin{figure}[h]
\includegraphics[width=0.9\columnwidth, clip]{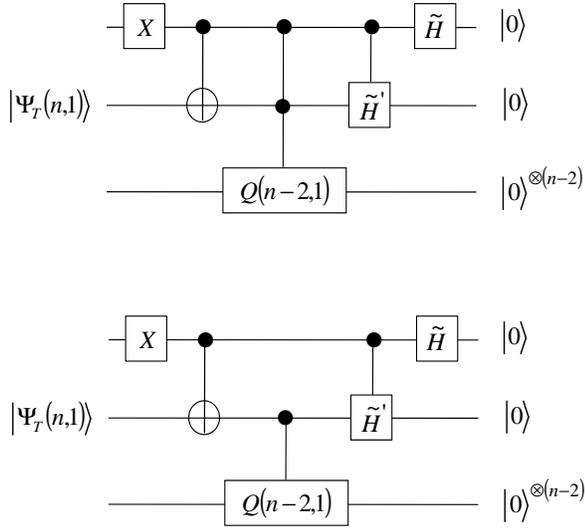}
\caption{Top quantum circuit for the recursive procedure to transform the
state $\vert \Psi _{T}(n,1)\rangle $ to the initial state $\vert 0\rangle
^{\otimes n}$. It has two one-fold control operations and one two-fold
operation in the middle. The bottom circuit shows its simplified version,
without two-fold operations.}
\label{re1}
\end{figure}

For the $m$-electron system, each of the three unitary operators in the
bottom line of the quantum circuit in Fig.~$6$ can be decomposed into the
circuit that has the same structure as that of $U(n,m)$. The circuit in Fig.~%
$6$ is divided in three parts. The two-fold controlled unitary operator in
the controlled operations in parts $I\!I$ and $I\!I\!I$ will appear when
they are expanded to one more qubit. Taking the state $\vert 0\rangle
^{\otimes n} $ as the input state, from right to left, following the
operation of the circuit, the basis vectors that appear in the state of the
qubits change in the same way as that of the one-electron system. These
twofold controlled unitary operators in parts I\!I and I\!I\!I of Fig.~$6$
are incompletely specified. For quantum circuit $U(n-1,m-1)$ that acts on
qubits $2,\cdots ,n, $ in part I of Fig.~$6$, the twofold controlled unitary
operators may appear when the circuit is decomposed one more step. From left
to right in Fig.~$6$, after the circuit in part I operates on the target
state $\vert \Psi _{T}(n,m)\rangle $, the target state is transformed to $%
c_{1}\vert 1\rangle U(n-1,m-1)\vert \Psi _{T}(n-1,m)\rangle +c_{0}\vert
0\rangle ^{\otimes n}$. This intermediate state can be divided in two
branches: the $\vert 0\rangle $ branch, $\vert 0\rangle ^{\otimes n},$ and
the $\vert 1\rangle $ branch, $\vert 1\rangle U(n-1,m-1)\vert \Psi
_{T}(n-1,m)\rangle $. For the $\vert 1\rangle $ branch, the operation of $%
U(n-1,m-1)$ that acts on qubits $2,\cdots ,n$, will be cancelled with the
controlled operation $C$-$U^{-1}(n-1,m-1)$ in part I\!I. For the $\vert
0\rangle $ branch of the state, the state on qubits $2,\cdots ,n,$ is $\vert
0\rangle ^{\otimes (n-1)}$, and it acts as the input state for the circuit $%
U(n-1,m-1)$ from right to left. Then the unitary operator $U(n-1,m-1)$ has
the same structure and input state as that of $U(n,m)$ we discussed above.
The twofold controlled unitary operators in $U(n-1,m-1)$ (part \negthinspace
I of Fig.~$6$) are also incompletely specified.
\begin{figure*}[tbp]
\includegraphics[width=2\columnwidth, clip]{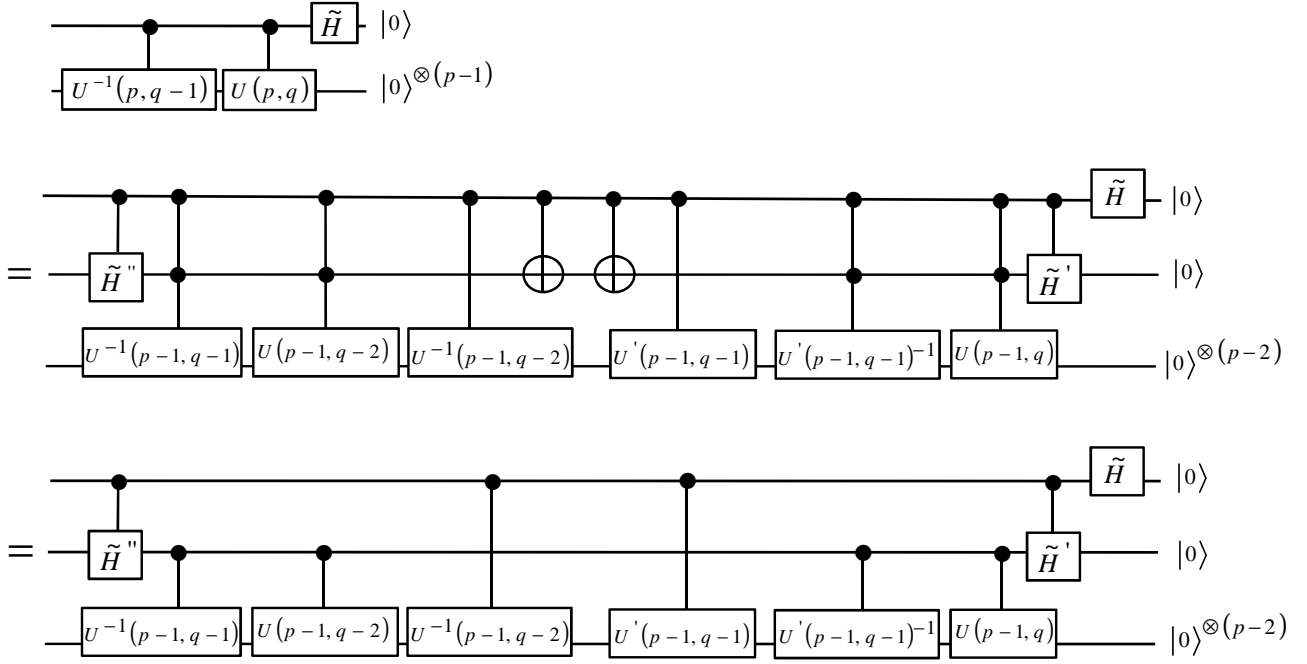}
\caption{Quantum circuit for two controlled unitary operators $C$-$U(p,q)$
and $C$-$U(p,q-1)$ and their simplified versions. The two unitary operators
in the top circuit are expanded in the second circuit, which is simplified
in the bottom circuit.}
\label{gu}
\end{figure*}

According to the analysis above, the unitary operator $U(n-1,m-1)$ has the
same structure as the circuit for $U(n,m)$, with state $\vert 0\rangle
^{\otimes (n-1)}$ as the input state. The twofold controlled unitary
operators in the controlled unitary operators in $U(n-1,m-1)$ and $U(n,m)$
are incompletely specified. A general form of the structure of the
controlled unitary operators is shown in Fig.~$8$. The twofold controlled
unitary operators in them are incompletely specified, and can be simplified
by just turning off the operation of the first control qubit as shown in
Fig.~$8$. Then all the twofold controlled gates in the circuit can be
simplified.

\section{Scaling of the algorithm}

\label{SCsec}

In this section, we analyze the scaling of the algorithm for state
preparation. The cost of a quantum algorithm is usually given by the number
of CNOT gates used. Here we count all the twofold controlled gates as
twoqubit gates since all twofold controlled gates can be simplified to
onefold controlled gates, as shown in Sec. V\!I\!I.

In the one-electron system, for states in the complete configuration space,
the scaling of the algorithm is derived as follows: let $N(n,1)$ denote the
total number of gates, including both CNOT and single-qubit gates, needed to
prepare the target state $|\Psi _{T}(n,1)\rangle $ from the initial state $%
|0\rangle ^{\otimes n}$. From the quantum circuit shown in Fig.~$7$, one can
see that there are four gates (a $\widetilde{H}$ gate is composed of three
gates: $C^{\dag }$, $C$, and NOT gates) on each line of the circuit, except
the last line, which has one gate. Therefore the total number of gates is $%
N(n,1)=4n-3$. The number CNOT gates can be derived as follows: denote $N_{%
\text{CNOT}}(n,1)$ as the number of CNOT gates. From Fig.~$7$, we have $N_{%
\text{CNOT}}(n,1)-N_{\text{CNOT}}(n-1,1)=2$(by adding one control qubit, one
obtains two more controlled gates, a CNOT gate and a $C$-$\widetilde{H}$
gate). Then keep this decomposition procedure until $n=2$, we obtain $N_{%
\text{CNOT}}(n,1)=2n-3$. This is the upper bound of the number of CNOT gates
for preparing a state in the one-electron system, since a target state in
the incomplete configuration space can be reduced to a state in the complete
configuration space with fewer qubits.

For the two-electron system, denote $N(n,2)$ as the number of gates needed
to prepare the target state $|\Psi _{T}(n,2)\rangle $from the $|0\rangle
^{\otimes n}$. Apply the decomposion procedure in Fig.~$6$ until $n=2$, and
using the result from the one-electron system that $N(n,1)=4n-3$, we obtain $%
N(n,2)=4n^{2}-10n+6$. The number of CNOT gates can be derived as follows:
let $N_{\text{CNOT}}(n,2)$ represent the number of CNOT gates needed to
prepare the target state $|\Psi _{T}(n,2)\rangle $. Apply the procedure in
Fig.~$6$ until $n=2$, and using the result from the one-electron system, we
obtain $N_{\text{CNOT}}(n,2)=2n^{2}-6n+4$.

The scaling for preparing a general state of the $m$-electron system is
derived as follows: denote $N(n,m)$ as the maximum number of gates,
including both CNOT and single-qubit gates, needed to prepare the state of
given values of $n$ and $m$. We should stress that all two-fold controlled
gates can be simplified to one-fold controlled gates~(see Sec.~IV). Looking
at Fig. $6$ and assuming that each block in it contains the maximum number
of gates possible for the size of the block, we have:
\begin{equation}
N(n,m)=2N(n-1,m-1)+N(n-1,m)+2,
\end{equation}%
The first order derivative of $N(n,m)$ with respect to $n$, where $N$ and $n$%
\ are integers and the derivative is taken accordingly, is:%
\begin{eqnarray}
\frac{dN(n,m)}{dn} &=&\frac{N(n,m)-N(n-1,m)}{n-(n-1)}  \notag \\
&=&2N(n-1,m-1)+2.
\end{eqnarray}%
Using Eq.~($34$), the second order derivative of $N(n,m)$ is:%
\begin{eqnarray}
\frac{d^{2}N(n,m)}{dn^{2}} &=&\frac{d}{dn}\left( \frac{dN(n,m)}{dn}\right)
\notag \\
&=&\frac{d}{dn}\left( 2N(n-1,m-1)\right)   \notag \\
&=&2^{2}N(n-2,m-2)+2^{2},
\end{eqnarray}%
After $(m-1)$ steps of taking the derivative of $N(n,m)$, we have:%
\begin{equation}
\frac{d^{m-1}N(n,m)}{dn^{m-1}}=2^{m-1}N(n-m+1,1)+2^{m-1}.
\end{equation}%
Using the scaling law given above for the one-electron case, i.e. $%
N(n-m+1,1)=4n-4m+1$, we find that:%
\begin{equation}
\frac{d^{m}N(n,m)}{dn^{m}}=2^{m-1}\text{ x }4=2^{m+1}.
\end{equation}%
We can now obtain the scaling behavior for preparing a general pure state of
the $m$-electron, $n$-spin-orbital system by integrating Eq.~($38$) $m$
times:%
\begin{equation}
N(n,m)\propto \frac{2^{m+1}}{m!}n^{m},
\end{equation}%
which scales \emph{polynomially} with $n$, the number of qubits in the
system. Since $N(n,m)$ is the maximum number of gates needed to prepare a
state with given values of $n$ and $m$, the scaling in Eq.~($39$) can be
thought of as the upper bound of the number of gates needed to prepare the
target state $|\Psi _{T}(n,m)\rangle $ from $|0\rangle ^{\otimes n}$. The
number of $C$-$\widetilde{H}$ gates is the majority of the circuit, e.g., as
shown in Fig. $8$; a $C$-$\widetilde{H}$ gate is composed of two
single-qubit gates, $C^{\dag }$ and $C$, and a CNOT gate. The number of CNOT
gates is comparable with the number single-qubit gates, both scaling as $%
O(2^{m}n^{m}/m!)$.

We now compare the scaling $O(2^{m}n^{m}/m!)$ with $O(2^{n})$, the number of
gates needed to prepare an arbitrary $n$-qubit state. We look for the
condition that makes $2^{m}n^{m}/m!$ smaller than $2^{n}$. Using the
Sterling approximation we have:
\begin{equation}
\frac{2^{m}}{m!}n^{m}\approx \frac{(2n)^{m}}{(\frac{m}{e})^{m}}=\left( \frac{%
2en}{m}\right) ^{m}.
\end{equation}%
Then the following equations need to be satisfied:%
\begin{equation}
\left( \frac{2en}{m}\right) ^{m}\leq 2^{n},
\end{equation}%
\begin{equation}
m\log _{2}\left( \frac{2en}{m}\right) \leq n,
\end{equation}%
\begin{equation}
m\log _{2}n<n.
\end{equation}%
One can see that if $m<n/\log _{2}n$, our algorithm is more efficient than
the algorithms mentioned in Refs.~\cite{shende} and \cite{mott}. Then we
compare the scaling of our algorithm with another algorithm introduced by
Ortiz \emph{et}~\emph{al.}~\cite{ortiz}, $O(N^{2}n^{2})$. Assuming $N\sim
\left(
\begin{array}{c}
n \\
m%
\end{array}%
\right) $, then looking for the condition that makes $2^{m}n^{m}/m!$ smaller
than $N^{2}n^{2}$, we have:
\begin{equation}
\frac{2^{m}}{m!}n^{m}<\left(
\begin{array}{c}
n \\
m%
\end{array}%
\right) ^{2}n^{2}\approx \left( \frac{n^{m}}{m^{m}}\right) ^{2}n^{2},
\end{equation}%
\begin{equation}
2^{m}<\left( \frac{n}{m}\right) ^{m}n^{2}.
\end{equation}%
One can see that as long as $n>2m$, our algorithm is more efficient than the
algorithm in Ref.~\cite{ortiz}. In quantum chemistry, usually $n\gg m$,
i.e., many more orbitals are needed to describe a given number of electrons.
The number of orbitals needed to describe a fixed number of electrons
depends on the accuracy of the calculation and the specific states that are
investigated. For example, to study the highly excited states of a molecular
system, more orbitals are needed than in study of the ground state. The
number of orbitals needed and the number of electrons are therefore
independent. So our algorithm gives an efficient way of preparing a general
state for accurate simulation of a wide range of molecular systems.

\section{Application to some molecular systems}

In this section, we apply the algorithm to two molecular systems: the
hydrogen molecule and the water molecule. We prepare the
multi-configurational self-consistent field~(MCSCF)~\cite{whf} wave function
of these two molecules. The MCSCF wave function is a linear combination of a
number of CSFs, where each CSF is a symmetry-adapted linear combination of
electron configurations.

The state function is a CSF, i.e., it is an eigen-function of the operators $%
\hat{L}^{2}$ and $\hat{S}^{2}$, where $\hat{L}$ and $\hat{S}$ are the
orbital angular momentum operator and the spin operator, respectively. In
quantum chemistry, one is usually most interested in the spin-states, i.e.,
the states that have certain spin multiplicity, of the molecular system. For
molecules that have space symmetry, the electronic states of the molecule
can be categorized into different irreducible representations of their point
group. One only needs to perform calculations on states that belong to a
certain irreducible representation. Considering these effects, the number of
configurations in the state function can be drastically below the size of
the Hilbert space. The state function is spanned only in a subspace of
dimension much smaller than $\left(
\begin{array}{c}
n \\
m%
\end{array}%
\right) $. Therefore many configurations do not appear in the decomposition
procedure shown in Eq.~($24$), where the state function is decomposed into a
$|0\rangle $ branch and a $|1\rangle $ branch. As a result, fewer gates are
needed to prepare the state than the conservative estimate given in Sec. V.
Note that the algorithm itself is not simplified in any way when considering
the symmetries of the molecule.

\subsection{Hydrogen molecule}

We use the MCSCF method with the cc-pVDZ basis set~\cite{dunning} to study
the electronic structure of the hydrogen molecule. This gives $20$ spin
orbitals. Two electrons are distributed on these orbitals with the
restriction of the spin multiplicity of the spin state. Considering the $%
D_{\infty h}$ symmetry of the molecule, the ground state of H$_{2}$ is $%
^{1}\Sigma _{g}^{+}$ ($^{1}\Sigma _{g}^{+}$ is the Mulliken symbol that is
used to label many-electron states; $\Sigma $, $g$, $+$, and the superscript
\textquotedblleft $1$" represents angular momentum, parity, symmetry with
respect to a vertical mirror plane perpendicular to the principal axis, and
the spin multiplicity).

For $20$ qubits, the dimension of the whole Hilbert space is $2^{20}\sim
10^{6}$; for a fixed number of electrons, the two-electron, $20$%
-spin-orbital system is of dimension $\left(
\begin{array}{c}
20 \\
2%
\end{array}%
\right) =190$; considering the spin multiplicity and the space symmetry, the
ground state MCSCF wave function is composed of at most $16$ electron
configurations. We can see that by considering the spin multiplicity and the
space symmetry, the state space is drastically reduced.

To prepare the state function in the full space of $\left(
\begin{array}{c}
20 \\
2%
\end{array}%
\right) $ using our state preparation algorithm, we need roughly $800$ CNOT
gates. To prepare the MCSCF wave function that is composed of $16$ electron
configurations, by performing a numerical calculation, we found that only $37
$ CNOT gates and $31$ single-qubit gates are needed in total. This number, $%
37$, is well below the conservative estimate of about $800$ CNOT gates
needed to prepare an arbitrary two-electron, $20$-spin-orbital state.

\subsection{Water molecule}

For another example, the water molecule H$_{2}$O, we use the MCSCF method
with the cc-pVDZ basis set~\cite{dunning}. For the ground state, considering
the $C_{2V}$ symmetry of the water molecule, the Hartree-Fock wave function
of the H$_{2}$O molecule is:
\begin{equation}
(1a_{1})^{2}(2a_{1})^{2}(1b_{2})^{2}(3a_{1})^{2}(1b_{1})^{2}\text{ .}
\label{babas7}
\end{equation}%
The ground state of H$_{2}$O is the $^{1}A_{1}$ state. We apply a complete
active space~(CAS) type MCSCF method in order to reduce the cost of the
calculation: the first two $a_{1}$ orbitals are frozen, the active space
consists of the $3a_{1}$-$6a_{1}$ orbitals, $1b_{1}$, $1b_{2}$ and $2b_{2}$
orbitals. So there are $14$ spin orbitals and $6$ electrons in the active
space.

For $14$ qubits, the dimension of the whole Hilbert space is $2^{14}\sim
10^{4}$; the six-electron, $14$-spin-orbital system is of dimension $\left(
\begin{array}{c}
14 \\
6%
\end{array}%
\right) =3003$. To prepare the state function in this space using our
algorithm, we need roughly $670000$ CNOT gates, which is larger than $%
2^{14}\sim 10^{4}$, the number of CNOT gates needed to prepare a state in
the full Hilbert space of $14$ qubits. Considering the spin multiplicity and
the space symmetry of the molecule, the ground state MCSCF wave function is
composed of $152$ electron configurations. To prepare the state function
using our state preparation algorithm, performing the numerical calculation,
we found that we need $1472$ CNOT gates and $1146$ single qubit gates. This
represents a considerable and remarkable reduction of over $3$ orders of
magnitude in the number of CNOT gates (from $\sim 670000$ to $\sim 1470$)
needed to prepare an arbitrary six-electron, $14$-spin-orbital state.

From these two examples, we can see that if the number of electrons is
comparable to the number of spin orbitals, the number of CNOT gates needed
to prepare the corresponding state increases very fast. The algorithm is no
longer efficient. However, in quantum chemistry, the number of orbitals
needed for simulation purposes is usually much larger than the number of
electrons. In the case of $n\gg m$, our algorithm provides an efficient way
in preparing states of molecular systems, as demonstrated in the example of
the hydrogen molecule above.

\section{conclusion}

\label{Consec}

In this paper, we present an efficient quantum algorithm for preparing a
pure molecular-system-like state. The simulated system lies in the Fock
space of a given number $m$ of electrons and $n$ spin orbitals on a quantum
computer. The state wave function is a configuration state function and is
entangled in general. In general this is to prepare a pure state in the
combinatorial space $\left(
\begin{array}{c}
n \\
m%
\end{array}%
\right) $ after the Jordan-Wigner transformation.

In our algorithm, instead of starting from an initial state and transforming
it to the target state, we start from the target state and transform it to
the initial state. Then by inverting the quantum circuit sequence, we obtain
the circuit for preparing the target state. A recursive procedure is
employed for solving this problem. The twofold controlled gates that appear
in each step of the recursive procedure can be simplified to one-fold
controlled gates by turning off the operation of the first control qubit,
since these gates are incompletely specified. We show that at most $%
O(2^{m+1}n^{m}/m!)$ gates, including both CNOT and single-qubit gates, are
needed to prepare a general state of the $m$-electron, $n$-spin-orbital
system, which scales \emph{polynomially} with $n$. The number of CNOT gates
scales as $O(2^{m}n^{m}/m!)$. This state preparation algorithm works for
systems of arbitrary number of electrons, but the number of CNOT gates
needed will increase exponentially if the number of electrons\ $m$ is
proportional to the number of spin orbitals $n.$

As examples, we have simulated our state preparation algorithm for the
hydrogen and water molecules. In these two specific cases we analyzed, using
the known symmetries of the molecules, we found that the number of CNOT
gates is reduced by up to $3$ orders of magnitude. This provides a
remarkable simplification to this type of problems.

\textit{Note added}.---In the final stages of preparing this paper, we
became aware of a paper that outlined a quantum algorithm for the
preparation of many-particle states on a lattice~\cite{guziksp}, in which
the state is prepared in the first-quantized representation, while ours is
in the second-quantized representation. We give an explicit and elaborate
solution to the problem and in a very different way.

\begin{acknowledgements}

FN acknowledges partial support from the National Security
Agency~(NSA), Laboratory for Physical Sciences~(LPS), (U.S.) Army
Research Office~(USARO), National Science Foundation~(NSF) under Grant
No.~EIA-0130383, and JSPS-RFBR under Contract No.~06-02-91200.
\end{acknowledgements}

\end{document}